**Temperature-dependent elastic properties of binary and multicomponent *high-entropy* refractory carbides**

D.G. Sangiovanni,[1*] F. Tasnádi,[1] T. Harrington,[2] K.S. Vecchio,[2] I.A. Abrikosov[1,3]

[1] Theoretical Physics Division, Department of Chemistry, Physics, and Biology (IFM), Linköping University, SE 581 83 Linköping, Sweden

[2] Department of NanoEngineering, University of California, San Diego, La Jolla, CA 92093, USA

[3] Materials Modeling and Development Laboratory, National University of Science and Technology "MISIS", 119049 Moscow, Russia

Available information concerning the elastic moduli of refractory carbides at temperatures (T) of relevance for practical applications is sparse and/or inconsistent. We carry out *ab initio* molecular dynamics (AIMD) simulations at T = 300, 600, 900, and 1200 K to determine the temperature-dependences of the elastic constants of rocksalt-structure (B1) TiC, ZrC, HfC, VC, and TaC compounds as well as multicomponent *high-entropy* carbides (Ti,Zr,Hf,Ta,W)C and (V,Nb,Ta,Mo,W)C. The second order elastic constants are calculated by least-square fitting of the analytical expressions of stress vs. strain relationships to simulation results obtained from three tensile and three shear deformation modes. Moreover, we employ sound velocity measurements to evaluate the bulk, shear, elastic moduli and Poisson's ratios of single-phase B1 (Ti,Zr,Hf,Ta,W)C and (V,Nb,Ta,Mo,W)C at ambient conditions. Our experimental results are in excellent agreement with the values obtained by AIMD simulations. In comparison with the predictions of previous *ab initio* calculations – where the extrapolation of finite-temperature elastic properties accounted for thermal expansion while neglecting intrinsic vibrational effects – AIMD simulations produce a softening of elastic moduli with T in closer agreement with experiments. Results of our simulations show that TaC is the system which exhibits the highest elastic resistances to both tensile and shear deformation up to 1200 K, and identify the high-entropy (V,Nb,Ta,Mo,W)C system as candidate for applications that require good ductility and toughness at room as well as elevated temperatures.

* Corresponding author: davide.sangiovanni@liu.se

## 1. Introduction

Elastic moduli and elastic stiffness constants of crystals have primary importance in materials design and materials discovery. For example, the search of novel solid crystal phases with potentially new properties includes verifying the Born von Kármán conditions of elastic stiffness constants for mechanical stability [1]. The temperature variation of the elastic anisotropy has a fundamental impact on the material's microstructure evolution and thus on observed age hardening mechanisms in ceramic coatings [2, 3]. Hardness and toughness of refractory ceramics are discussed in terms of the residual stress in the material, which is evaluated through the knowledge of elastic stiffness constants. Indeed, *ab initio* investigations have shown that the predicted elastic constants may vary up to ≈20% depending on deformation path, strain range, and order of the polynomial used to fit energy vs. strain curves [4]. In regard to the identification of materials with superior mechanical or structural performance, the observed trends in hardness [5, 6], strength [7, 8], toughness [9-11] and ductility [7, 8, 12, 13] are often correlated to trends in *elastic* constants of the materials.

The elastic constants of solid crystal phases can be efficiently screened by high-throughput *ab initio* calculations at 0 kelvin [14]. In contrast, realistic modeling and evaluation of mechanical properties as hardness and toughness, which would require simulation boxes with volumes of the order $\approx 10^3$–$10^5$ $nm^3$ [15], is currently unfeasible for first-principles methods. For this reason, *ab initio* calculated polycrystalline elastic moduli, such as shear modulus and Poisson ratio, are widely employed to design empirical indicators of hardness and toughness of compounds and alloys [9, 16-20] which, combined with the progress in machine-learning approaches [20-22], can lead to an increasingly more rapid identification of solids with enhanced mechanical performance.

In the case of refractory carbide, nitride, and boride ceramics, properties as hardness and toughness are of paramount importance, especially for high-temperature applications. Despite this, the elastic constants of these refractory ceramics have been primarily evaluated at 0 K. *Ab initio* estimations of elastic constants at finite temperatures are often done using static calculations at cell volumes determined in the quasiharmonic approximation for the thermal expansion (see, e.g., [23-27]). However, explicit inclusion of lattice vibrations in theoretical modeling is shown to result in significant differences in second-order elastic moduli and a change of elastic anisotropy of materials at moderate or high temperatures [28, 29]. *Ab initio* molecular dynamics (AIMD) simulations, which inherently reproduce anharmonic vibrational effects, provide more reliable predictions of elastic constants at elevated temperature [29-31].

Recently, we have reported on an *ab initio* molecular dynamics (AIMD) approach for computing stress/strain curves of single-crystal B1 ceramics subject to tensile and shear deformation at any temperature of interest [32-34]. The simulations employ deformations well beyond the elastic limit conventionally used to calculate the elastic constants [28-31], and allow evaluating mechanical properties including, e.g., ideal strengths, moduli of resilience, toughness, and elongation at fracture. The method detailed in Refs. [32-34] has been used to produce the full stress/strain curves in rocksalt-structure (B1) transition-metal carbides (TMC) at temperatures $300 \leq T \leq 1200$ K [35]. In order to determine the most essential parameters that characterize the mechanical response of the studied systems, we here exploit the stress/strain relationships evaluated in the elastic regime [35] to extract the second-order elastic constants (SOEC) of B1 TiC, ZrC, HfC, VC, and TaC, as well as high-entropy (Ti,Zr,Hf,Ta,W)C and (V,Nb,Ta,Mo,W)C. The fact that our approach utilizes six different deformation pathways (see **Section 2.1**) provides SOEC with reduced numerical dependence on the specific choice of strain matrixes and strain intervals [4]. For the two high-entropy carbides (HEC), synthesized as single-phase B1 materials [36-38], we also evaluate the room-temperature elastic moduli from sound velocity measurements that allow us to verify the reliability of AIMD results. We note that the (Ti,Zr,Hf,Ta,W)C system exhibits particularly high hardness [36], while (V,Nb,Ta,Mo,W)C is of interest because of its relatively high valence electron concentration, which may improve ductility [9, 39].

# 2. Methods

## 2.1. AIMD simulations

In this work, we employ tensile and shear stress/strain data within the elastic (prior to yielding) material's responses obtained from AIMD simulations performed at T = 300, 600, 900, and 1200 K [40] and evaluate the SOEC of B1 TMC as a function of temperature (T). The entire (elastic and plastic) stress vs. strain curves calculated for TMC subject to tensile and shear deformation will be reported elsewhere [35]. The procedure for AIMD stress/strain calculations is detailed in Refs. [32-34]. The B1 TMC supercells used in our simulations have three different crystallographic orientations (**Fig. 1**), which are convenient for modeling elongation orthogonal to (001), (110), and (111) (i.e., surface terminations of lowest formation energies [41]) and shearing on the primary slip systems of B1-structure ceramics, namely, $\{001\}\langle 1\bar{1}0\rangle$, $\{110\}\langle 1\bar{1}0\rangle$, and $\{111\}\langle 1\bar{1}0\rangle$ [42-45]. The stress vs. strain relationships are fitted with $2^{nd}$-order polynomials. Then, the second-order elastic response of the crystal is calculated from the slope of the polynomial in the limit of vanishingly small strain. The approach is analogous to fitting the energy-density vs. strain curves with $3^{rd}$-order polynomials, as illustrated, for example, in Ref. [4].

AIMD simulations are carried out with VASP code [46], which implements the projector augmented-wave method [47]. The Perdew-Burke-Ernzerhof (PBE) electronic exchange and correlation functionals of the generalized gradient approximation (GGA) [48], Γ-point sampling of the Brillouin zone, and planewave cutoff energy of 300 eV are used in all simulations. The ionic equations of motion are integrated at timesteps of 1 fs, using $10^{-5}$ eV/supercell convergence criteria for the total energy during both system equilibration and mechanical testing. Prior to modeling tensile and shear deformation, the supercell equilibrium structural parameters are optimized at each investigated temperature via an iterative NPT [49] simulation procedure, which employs the Parrinello-Rahman barostat [50] and the Langevin thermostat [51]. Subsequently, NVT [52] sampling of the configurational space is used to equilibrate the unstrained structures for 3–5 additional ps using the Nosé-Hoover thermostat [51]. In these simulations, we ensure that the time-averaged $|\sigma_{xx}|$, $|\sigma_{yy}|$, and $|\sigma_{zz}|$ stress components are $\lesssim$ 0.3 GPa. The structural parameters determined for unstrained (equilibrium) supercells at different temperatures allow us to evaluate lattice constants and average linear thermal expansion coefficients of B1 TMC.

$\langle 001\rangle$, $\langle 110\rangle$, $\langle 111\rangle$ tensile and $\{001\}\langle 1\bar{1}0\rangle$, $\{110\}\langle 1\bar{1}0\rangle$, and $\{111\}\langle 1\bar{1}0\rangle$ shear deformations of B1 stoichiometric TMC are increased at sequential increments of strain $\delta$=2%, see details in Refs. [32-34]. At each strain, the crystals are equilibrated for (at least) two ps via canonical sampling. All simulation structures contain 576 atoms: 24 atomic layers with 24 atoms per layer. In HEC, the metals are stochastically distributed on the cation sublattice. The values of total stress (including kinetic contributions) values that we report are averaged over 500 equilibrated AIMD configurations.

## 2.2 Determination of SOEC from AIMD

The six different deformations [three tensile: $\langle 001\rangle$, $\langle 110\rangle$, and $\langle 111\rangle$, with supercell elongation in the vertical z direction; three shearing: $\{001\}\langle 1\bar{1}0\rangle$, $\{110\}\langle 1\bar{1}0\rangle$, and $\{111\}\langle 1\bar{1}0\rangle$ with trigonal strain of supercells within the xz plane (see supercells in **Fig. 1**)] employed in our AIMD simulations, result in 36 independent stress values for each applied tensile or shear strain ($\delta$). $\delta$ is the strain increment in either deformation. Of these, thirteen stress components – namely $\sigma_{xx}^{\langle 001\rangle}$, $\sigma_{yy}^{\langle 001\rangle}$, $\sigma_{zz}^{\langle 001\rangle}$, $\sigma_{xx}^{\langle 110\rangle}$, $\sigma_{yy}^{\langle 110\rangle}$, $\sigma_{zz}^{\langle 110\rangle}$, $\sigma_{xx}^{\langle 111\rangle}$, $\sigma_{yy}^{\langle 111\rangle}$, $\sigma_{zz}^{\langle 111\rangle}$, $\sigma_{xz}^{\{001\}\langle 1\bar{1}0\rangle}$, $\sigma_{xz}^{\{110\}\langle 1\bar{1}0\rangle}$, $\sigma_{xy}^{\{111\}\langle 1\bar{1}0\rangle}$, and $\sigma_{xz}^{\{111\}\langle 1\bar{1}0\rangle}$ – are of relevance to our analysis because their derivative $\left[\frac{\partial\sigma}{\partial\delta}\right]_{\delta=0}$ corresponds to a linear combination of elastic constants [53]. Least-squares fitting is used to determine 2nd order polynomials which best describe the dependences of each of the stress components listed above as a function of $\delta$. The polynomial's first derivatives at 0 strain are gathered in a vector. Then, the canonical elastic stiffness tensor $C_{ij}$ is transformed into the coordinate systems of our supercells (see **Fig. 1**). The elastic constant combinations corresponding to the stresses employed in our analysis are summarized in **Table 1**. The thirteen stress vs. strain relationships define a set of linear equations, which can be written in a matrix expression:

$$\begin{pmatrix} 0 & 1 & 0 \\ 0 & 1 & 0 \\ 1 & 0 & 0 \\ 1/2 & 1/2 & -1 \\ 0 & 1 & 0 \\ 1/2 & 1/2 & 1 \\ 1/3 & 2/3 & -2/3 \\ 1/3 & 2/3 & -2/3 \\ 1/3 & 2/3 & 4/3 \\ 0 & 0 & 1 \\ 1/2 & -1/2 & 0 \\ \sqrt{2}/6 & -\sqrt{2}/6 & -\sqrt{2}/3 \\ 1/3 & -1/3 & 1/3 \end{pmatrix} \begin{pmatrix} C_{11} \\ C_{12} \\ C_{44} \end{pmatrix} = \left[\frac{\partial}{\partial \delta}\right]_{\delta=0} \begin{pmatrix} \sigma_{xx}^{\langle 001 \rangle} \\ \sigma_{yy}^{\langle 001 \rangle} \\ \sigma_{zz}^{\langle 001 \rangle} \\ \sigma_{xx}^{\langle 110 \rangle} \\ \sigma_{yy}^{\langle 110 \rangle} \\ \sigma_{zz}^{\langle 110 \rangle} \\ \sigma_{xx}^{\langle 111 \rangle} \\ \sigma_{yy}^{\langle 111 \rangle} \\ \sigma_{zz}^{\langle 111 \rangle} \\ \sigma_{xz}^{\{001\}\langle 1\bar{1}0 \rangle} \\ \sigma_{xz}^{\{110\}\langle 1\bar{1}0 \rangle} \\ \sigma_{xy}^{\{111\}\langle 1\bar{1}0 \rangle} \\ \sigma_{xz}^{\{111\}\langle 1\bar{1}0 \rangle} \end{pmatrix}. \tag{1}$$

Finally, the cubic SOEC can be derived by solving the overdetermined system of equations (**Eq. (1)**) by a least square fitting approach (multiple linear regression). The procedure allows us reducing the numerical uncertainties in calculated SOEC values due to a specific choice of strain matrix or strain range [4].

The elastic constants $C_{11}$, $C_{12}$, $C_{44}$ obtained from **Eq. (1)** are used to compute other elastic properties. Specifically, we derive bulk moduli B, polycrystalline shear G and Young's E moduli according to the Voigt ($G_V$ and $E_V$), Reuss ($G_R$ and $E_R$), and Hill ($G_H$ and $E_H$) approximations (formulas in [30]), directional Young's moduli $E_{001}$, $E_{110}$, and $E_{111}$ (expressions in [54]), and elastic shear resistances $G_{110} = (C_{11}–C_{12})/2$, $G_{111u} = (C_{11}+C_{12}+C_{44})/3$, and $G_{111r} = 3C_{44}(C_{11}–C_{12})/(4C_{44}+C_{11}+C_{12})$ resolved onto $\{110\}\langle 1\bar{1}0\rangle$ and $\{111\}\langle 1\bar{1}0\rangle$ slip systems. The subscripts u and r are used to label unrelaxed (constant strain) and relaxed (constant stress) $G_{111}$ shear moduli. In this work, we focus on the $G_{111u}$ rather than $G_{111r}$ moduli, because our simulations are performed with constant strain boundary conditions. The Poisson's ratios ν are computed as $\nu = ½ [1–E_H/(3B)]$.

### 2.3 Material synthesis and elastic constants measurements

For fabricating the high entropy carbides used in this study, binary carbide powders were weighted out in 24 gram quantities to achieve an equiatomic mixture of the 5 metallic species, additional graphite was added to yield the stoichiometric monocarbide ratios when $Mo_2C$ was included. The binary carbides included: TiC, ZrC, HfC, VC, TaC, $Mo_2C$ and WC, all obtained from Alfa Aesar, USA. The powders were blended and hand mixed to achieve the following bulk compositions: (Ti,Zr,Hf,Ta,W)C and (V,Nb,Ta,Mo,W)C, followed by high energy ball milling in 30 minute increments, for a total of 120 minutes, using 10-minute cooling periods between each step. The powder samples were then compacted and sintered using current and pressure assisted densification (CAPAD) in a well-controlled vacuum environment. The samples were sintered using 20mm diameter graphite dies and plungers. The initial vacuum condition was 20 mTorr, and upon heating, whenever the vacuum decreased to 100 mTorr, the heating profile was paused to allow the vacuum to return to the 20 mTorr level. The heating and pressure profile used in the sintering of both materials was as follows: using an initial pressure of 5 MPa, the samples were heated at 100°C per minute to 2200°C, then held at temperature for 9 minutes, before increasing the pressure to 80 MPa over one minute, then holding at this pressure and 2200°C for 15 minutes. The samples were subsequently cooled to room temperature in vacuum under 5 MPa pressure. Each sample was ground on both circular faces to remove any excess graphite from the die setup, then polished using standard metallurgical techniques to a final polish of 0.04 µm colloidal silica. The carbon-to-metal ratios assessed for our HEC specimens are 0.90±0.03 [55].

To obtain the elastic moduli, both longitudinal and shear, the method outlines in ASTM E494-15 (Standard Practice for Measuring Ultrasonic Velocity in Materials) were followed, using the 20 mm diameter and ~3 mm thick sintered high entropy carbide samples. The samples were placed on an acoustically-isolated substrate for wave speed measurements using separate piezoelectric transducers for longitudinal and shear velocities. A total of 4 separate measurements, in each velocity direction, were obtained in order to compute the respective wave speeds from which to compute the different elastic properties. To compute the various elastic constants, both the respective wave speeds and the sample density must be measured, with the latter measured by Archimedes method. The following equations were used to compute the elastic properties, which assume isotropic elastic responses. For Poisson's ratio: $v = (V_L^2 - 2V_S^2)/[2(V_L^2 - V_S^2)]$; for Young's modulus: $E = 2\,V_S^2\,\rho\,(1+v)$; for shear modulus: $G = V_S^2\,\rho$; and for bulk modulus: $B = E/[3(1-2v)]$, where $\rho$ is density, $V_L$ is the measured longitudinal wave speed and $V_S$ is the measured shear wave speed.

## 3. Results and discussion

The calculated and experimental lattice parameters and average linear thermal expansion coefficients $\alpha$ of B1 TiC, ZrC, HfC, VC, TaC, (Ti,Zr,Hf,Ta,W)C, and (V,Nb,Ta,Mo,W)C at temperatures ranging from 0 to 1200 K are presented in **Table 2 (AIMD data listed in bold)**. Previous 0-K *ab initio* calculations based on GGA slightly overestimate (by less than 1%) the experimental lattice parameters of B1 TMC (see **Table 2**). The discrepancy is due to known limitations of GGA functionals [56], as well as to the fact that the synthesized transition-metal carbides typically contain anion vacancies, which reduce equilibrium volumes [57, 58]. At deviance with *ab initio* results at 0 K, our AIMD simulations yield lattice constants that are systematically ≈0.5% smaller than experimental values measured at 300 K. This is likely due to the relatively lower accuracy of present simulations (details in **Section 2.1**) in comparison with static *ab initio* calculations, for which the structural parameters are converged with respect to cutoff energies and k-point mesh thicknesses at low computational efforts. Overall, our AIMD results closely match with the lattice constants and thermal expansion coefficients determined by experiments at different temperatures (**Table 2**).

The SOEC of refractory carbides, nitrides, and borides have been extensively investigated via first-principles energy-minimization methods at 0 kelvin [4, 16-18, 23, 59-61]. In contrast, information concerning finite-temperature elastic properties is rather sparse for these classes of materials. Finite-temperature estimations of SOEC can be efficiently obtained using static *ab initio* calculations that employ the quasiharmonic approximation (QHA) [23, 27, 62] or volume-scaling methods [28, 31]. These methods account for lattice thermal expansion while neglect anharmonic vibrational effects. The latter effects can yield significant corrections to SOEC values calculated at high temperatures using static *ab initio* approaches, as demonstrated for the cases of B1 TiN, CrN, and (Ti,Al)N compounds [28, 31]. A more detailed comparison among AIMD, static *ab initio*, and experimental results of elastic properties as a function of temperature can be found below, toward the end of **Section 3**.

To the authors' knowledge, very few studies based on AIMD simulations – which inherently describe lattice vibrations – focused on direct evaluation of the elastic properties of refractory ceramics [30-33, 63]. The elastic properties of B1 TiC, ZrC, HfC, VC, TaC, (Ti,Zr,Hf,Ta,W)C, and (V,Nb,Ta,Mo,W)C determined by AIMD simulations between 300 and 1200 K are listed in **bold in Tables 3–9**. For comparison, the tables also contain data reported in other theoretical and experimental works. Previous studies mainly focused on synthesis and investigation of group-IV carbides. Measurements and *ab initio* results for the properties of VC and TaC are sparser. In addition, the information available for the B1 carbides of vanadium and tantalum show considerable incongruences. For example, the Young's moduli measured at 300 K are scattered in the range 268 – 420 GPa for vanadium carbides and 241 – 722 GPa for tantalum carbides [36]. To some extent, such large deviations can be attributed to variations in metal/carbon stoichiometries and different porosity of the samples. In the case of HEC, a new class of refractory ceramics, the information available in the literature is limited to properties calculated at 0 K or determined by experiments at room temperature [36, 37].

Overall, our AIMD results closely match with experimental [64] and *ab initio* values retrievable from the literature (**Tables 3–9**). For example, the temperature dependence of the Young's moduli E of B1 zirconium carbides, previously determined by experiments [65-67] and other *ab initio* calculations based on the QHA [27], agrees with the results of AIMD simulations (**Table 4**). The variation of E vs. T shows that the elastic modulus of ZrC decreases at a quasi-linear decay rate $(1/E_{300K})\times(dE/dT) \approx -1.1\pm0.2 \times 10^{-4}$ $K^{-1}$ for temperatures increasing from 300 to 1200 K. Note that the decay coefficient varies with the sample stoichiometry [67]. A decrease in elastic stiffness vs. T is expected for crystal phases that are dynamically stable from 0 K to melting points. The monotonic reduction in SOEC with increasing temperature [68-70] is also reflected by a corresponding decrease in mechanical strength and hardness [42, 71]. Deviations from the typical behavior of SOEC vs. T – see, e.g., temperature-induced increase in the $C_{44}$ of some metals [72, 73] – may arise from strain- or thermally-induced modifications of electronic-structures.

All TMC modelled in this work display expected trends in $C_{11}$ and $C_{44}$ vs. temperature. Both elastic constants decrease monotonically from 300 to 1200 K (**Fig. 2a and 2c**). Conversely, the values of the $C_{12}$ elastic constants appear to be less affected by temperature. Indeed, the $C_{12}$ calculated for TaC, HfC, and (Ti,Zr,Hf,Ta,W)C remains nearly unvaried (within statistical uncertainty) with T (**Fig. 2b**). The trends that we obtain for all other carbides indicate slow $C_{12}$ reductions with increasing temperature (**Fig. 2b**). The diminution in $C_{11}$ and $C_{44}$ implies that the moduli B, G, and E decrease as a function of temperature for all materials considered here (**Fig. 2e-f**).

Consistent with the temperature dependence observed for the polycrystalline moduli G and E, the directional elastic $E_{001}$, $E_{110}$, $E_{111}$ and shear $G_{001}$, $G_{110}$, $G_{111u}$ moduli of the investigated carbide systems decrease with increasing T (**Figs. 2c, 3a-e**). The moduli $E_{hkl}$ and $G_{hkl}$ of each material strictly maintain a mutual relationship [$E_{001} > E_{110} > E_{111}$] and [$G_{110} > G_{111u} > G_{001}$ ($\equiv C_{44}$)] from 300 and up to 1200 K. This can be seen by gathering $E_{hkl}$ and $G_{hkl}$ data from **Figs. 2c, 3a-e**. Given that $E_{001} > E_{110} > E_{111}$, it is concluded that the TMC systems retain similar directionality in bond stiffnesses at all investigated temperatures; this is also reflected by the calculated trends in coefficients A (**Tables 3–9**) which, overall, indicate a modest increase in TMC elastic isotropy between 300 and 1200 K.

The fact that $G_{110}$ – elastic resistance to {110} plane shearing along $\langle 1\bar{1}0 \rangle$ – and $G_{001}$ – elastic resistance to {001} plane shearing along $\langle 1\bar{1}0 \rangle$ – are systematically the highest and lowest shear moduli may (naively) appear inconsistent with experimental findings for plastic deformation mechanisms in B1 carbides. In this regard, several experimental studies showed that $\{110\}\langle 1\bar{1}0 \rangle$ slip is the most active in transition metal carbides at low temperature, whereas lattice slip on {111} planes becomes generally operative at moderate or high temperatures [42, 43, 74]. In contrast, $\{001\}\langle 1\bar{1}0 \rangle$ glide in carbides has been experimentally observed in very few cases [44, 45]. Nonetheless, a stiffer elastic response to change of shape (quantified by the shear moduli) does not necessarily imply that the shear strength (resistance to slip) is larger. For example, B1 $VN_{0.8}$ exhibits greater elastic shear stiffness, but much lower shear strength than B1 VN [32]. Analogously, although TaC displays the highest shear moduli $G_{111u}$ among all TMC (**Fig. 3b**), $\{111\}\langle 1\bar{1}0\rangle$ plastic deformation of B1 TaC crystals can be observed already at room temperature [74].

At relatively low temperatures, mechanical strength and plastic deformation in B1-structure ceramics is primarily controlled by dislocation mobilities [43-45, 74]. The mobility is, in turn, dictated by the core structure of the line defect. An atomic-level determination of dislocation core structures in ceramics can be a formidable task. The problem is further complicated by the fact that dislocation cores may exhibit different polymorph variants [75, 76] and be affected by lattice stoichiometry [77]. For these reasons, several empirical criteria have been proposed for rapidly predicting qualitative trends in ductility and strength among different materials. For example, the criterion proposed by Pugh [12] has a broad applicability because it bases its predictions on the ratio between polycrystalline moduli G and B. Pettifor's criterion [13], on another hand, is useful to estimate trends in ductility among crystals with cubic symmetry. Remarkably, the use of these phenomenological criteria [9, 17] has led to the discovery of hard and inherently tough ceramics [10].

Recently, a generalization of Pugh's and Pettifor's criteria [7], in which the Cauchy's pressure $C_{12}$–$C_{44}$ is normalized with the modulus E, has been proposed for reproducing experimental trends in ductility vs. strength across different classes of solids. The empirical criterion of Niu *et al.* [7] suggests that strength and ductility are necessarily mutually exclusive properties in crystals: high G/B (high strength) implies low $(C_{12}$–$C_{44})$/E (poor ductility ≡ brittleness), and vice-versa. The map in ductility vs. strength determined for B1 TMC is illustrated in **Fig. 3f**, where G/B vs. $(C_{12}$–$C_{44})$/E are plotted for each investigated temperature. The AIMD results in **Tables 3, 4, 5, and 8** show that the G/B and $(C_{12}$–$C_{44})$/E values calculated for the three Group-IVB carbides, as well as B1 (Ti,Zr,Hf,Ta,W)C, remain approximately constant with temperature. We note that B1 HfC is the crystal phase characterized by the highest G/B values, and lowest $(C_{12}$–$C_{44})$/E, from 300 to 1200 K (**Fig. 3f**). Conversely, B1 VC, TaC, and (V,Nb,Ta,Mo,W)C are predicted to become more ductile (and less strong) with increasing temperature (see arrow in **Fig. 3f**).

Previous *ab initio* studies conducted on refractory B1-structure ceramics indicated that an increased valence electron concentration (VEC) enhances the material's ductility [9, 16-18]. Overall, the trends illustrated in **Fig. 3f** are aligned with such empirical rule. Indeed, a plot of $[G/B]_{VEC}$ vs. $[(C_{12}–C_{44})/E]_{VEC}$ – where "$[\ldots]_{VEC}$" denotes a property-average for an isoelectronic class of TMC – predicts that Group-IVB carbides (VEC=8.0 $e^{-}$/f.u.) are the strongest, but also the most brittle systems, whereas (V,Nb,Ta,Mo,W)C (VEC=9.4 $e^{-}$/f.u.) is the material which, by far, exhibits the best ductility (inset in **Fig. 3f**). The strength G/B and ductility $(C_{12}–C_{44})/E$ indicators calculated for the high entropy (Ti,Zr,Hf,Ta,W)C system are intermediate to the values obtained for binary Group-IVB and Group-VB TMC (inset in **Fig. 3f**). This is consistent with the VEC rule, because the HEC's electron concentration (8.6 $e^{-}$/f.u.) is between those of Group-IVB (8.0 $e^{-}$/f.u.) and Group-VB (9.0 $e^{-}$/f.u.) carbides.

The theoretical prediction that TaC and VC should be more ductile than TiC, ZrC, and HfC is consistent with the fact that $\{111\}\langle 1\bar{1}0\rangle$ slip in Group-VB carbides is activated at lower temperatures than in Group-IVB carbides [42]. The operativity of lattice slip on {111} crystallographic planes is of crucial importance for ductility in B1-structure ceramics, because it provides a sufficient number of independent pathways to induce plastic deformation [78]. In agreement with the conclusions of Ref. [42], other experiments have shown that the concentration of intrinsic {111} stacking faults in B1 TaC is much greater than in, e.g., B1 HfC [79]. The observation has been rationalized by *ab initio* results, which indicated that these extended defects are metastable in Group-VB, whereas they are unstable in Group-IVB carbides (see [79] with supplemental material). The energetically-favored formation of {111} stacking faults facilitates, in turn, plastic deformation via $\{111\}\langle 11\bar{2}\rangle$ synchro-shear mechanisms [79].

The results of our AIMD simulations demonstrate that TaC is the system of highest elastic resistances to tensile ($C_{11}$, E, $E_{001}$, $E_{110}$, $E_{111}$) as well as shear ($C_{44}$, G, $G_{110}$, $G_{111u}$) deformation at all investigated temperatures, **Figs. 2 and 3**. VC is the crystal with second highest tensile and shear elastic stiffnesses (with a few exceptions for T ≈1200 K). Conversely, we find that B1 ZrC generally displays the lowest values of G and E elastic stiffnesses. With regard to high entropy alloys, **Fig. 2a** shows that the $C_{11}$ elastic constants of (Ti,Zr,Hf,Ta,W)C and (V,Nb,Ta,Mo,W)C are nearly identical, and with values intermediate to those of Group-IV and Group-V carbides at the temperatures probed in the present work. Our AIMD results also demonstrate that (V,Nb,Ta,Mo,W)C possesses the lowest $C_{44}$ and highest $C_{12}$ values for $300 \leq T \leq 1200$ K (**Fig. 2c**). As discussed above and illustrated in **Fig. 3f**, this is reflected by its highest predicted ductility (empirical indicator $(C_{12}–C_{44})/E$ [7]).

Among numerous multicomponent carbide alloys synthesized during recent years [36-38, 80-83], here we focus on B1 (Ti,Zr,Hf,Ta,W)C and B1 (V,Nb,Ta,Mo,W)C due to expectations for particularly high hardness and toughness, respectively. The VEC (8.6 $e^{-}$/f.u.) of (Ti,Zr,Hf,Ta,W)C is close to the value (≈8.4 $e^{-}$/f.u.) that maximizes hardness in B1-structure carbonitrides [5]. The electronic mechanism for increased hardness in the single-crystal B1 phase finds its origin in fully occupied shear-resistant metal(d)–non-metal(p) vs. empty shear-sensitive metal(d)–metal(d) electronic states. In the case of (V,Nb,Ta,Mo,W)C (9.4 $e^{-}$/f.u.), we envisage an energetically-favored formation of {111} stacking faults (during synthesis or upon loading), as previously demonstrated for B1-structure carbides and nitrides of similar VEC (≈9.5 $e^{-}$/f.u). [84, 85]. Hindered dislocation motion across the faults provides high hardness and mechanical strength [84, 85]. (V,Nb,Ta,Mo,W)C does indeed exhibit rather high hardness values (27±3 GPa [37]). Concurrently, a substantial amount of Mo and W in the host B1 (V,Nb,Ta,Mo,W)C lattice should promote slip on {111} lattice planes [86], which fulfills the criteria for plasticity in B1-structure crystals [78]. $\{111\}\langle 11\bar{2}\rangle$ synchro-shear, activated upon loading in alloy domains characterized by locally higher concentrations of Group-VIB metals, produces {111} stacking faults or twins in the B1 lattice (note that near-stoichiometric $MoC_x$ and $WC_x$ are thermodynamically-inclined to form the hexagonal WC structure [87, 88] with A-B-A-B stacking sequence along [0001]). Relatively facile $\{111\}\langle 11\bar{2}\rangle$ deformation provides a transformation pathway for stress dissipation and thus enhances toughness.

The elastic moduli of the two HEC systems investigated in this work have been previously assessed by experiments and *ab initio* calculations [36]. Our AIMD results for B, G, E, and ν of the HECs at 300 K closely match (deviations < 5%) with first-principles elastic properties at 0 K [36] and are also in good agreement with previous experimental data [36] (see **Tables 8 and 9**). Our present measurements of acoustic wave speeds in the HECs return elastic stiffnesses and Poisson's ratios (see **(values)** in **Tables 8 and 9**) that are in quantitative agreement with AIMD predictions. Our AIMD vs. (experimental) results are (*i*) bulk moduli $B^{(Ti,Zr,Hf,Ta,W)C}$ = 288 (273±17) GPa, $B^{(V,Nb,Ta,Mo,W)C}$ = 311 (269±12) GPa; (*ii*) shear moduli $G^{(Ti,Zr,Hf,Ta,W)C}$ = 194 (193±4) GPa, $G^{(V,Nb,Ta,Mo,W)C}$ = 174 (173±4) GPa; (*iii*) Young's moduli $E^{(Ti,Zr,Hf,Ta,W)C}$ = 475 (468±3) GPa, $E^{(V,Nb,Ta,Mo,W)C}$ = 440 (426±5) GPa; (*iv*) Poisson's ratios $\nu^{(Ti,Zr,Hf,Ta,W)C}$ = 22 (21±2) %, $\nu^{(V,Nb,Ta,Mo,W)C}$ = 26 (24±1) %.

Before conclusions, we offer a comparison of different computational techniques for the evaluation of finite-temperature elastic properties. As anticipated at the beginning of this section, finite-temperature elastic properties of some refractory nitrides and carbides have been calculated via the QHA [23, 27, 62] or volume-scaling methods [28, 31] in the framework of *static* first-principles techniques. These, however, generally produce a slower decay in elastic stiffnesses vs. T than methods which account for intrinsic anharmonic effects, as demonstrated for B1 $Ti_{1-x}Al_xN$ [28] and CrN [31] refractory compounds. Here we extend the analyses of Refs. [28, 31] by showing trends in elastic constants vs. T obtained by AIMD simulations, experiments, and static *ab initio* calculations.

**Fig. 4** illustrates the temperature dependences of the elastic shear stiffnesses of **(a)** TiC, **(b)** ZrC, and **(c)** TiN. Note that B1 TiN is included in our analysis because of the sparseness of temperature-dependent elastic properties for refractory carbides. A detailed description of the elastic properties of TiN and other nitrides will be provided elsewhere [35]. In all three cases presented in **Fig. 4**, one can see that the explicit treatment of lattice vibrations – inherent to AIMD simulations – yields a decrease of shear-stiffness vs. T in closer agreement with experimental trends. Conversely, static *ab initio* and AIMD results of other elastic properties are in apparently similar agreement with experimental trends (compare, e.g., $C_{11}$ and $C_{12}$ vs. T for TiC, **Table 3**, and B and E vs. T for ZrC, **Table 4**). This suggests that the inclusion of vibrational effects is of particular importance to achieve accurate temperature-dependent elastic shear stiffnesses $C_{44}$, which, in turn, strongly affect the values of polycrystalline shear moduli G.

## 4. Conclusions

We employ AIMD simulations and sound velocity measurements to provide an accurate database of elastic properties of B1-structure refractory compound and multicomponent carbides from 300 to 1200 K. Exception made for the $C_{12}$, our results show that the elastic constants of all carbides decrease monotonically with increasing temperature. B1 TaC is the material which displays the highest values of $C_{11}$, $C_{44}$, B, G, and E moduli up to 1200 K. A comparison of the results of present AIMD simulations, previous static *ab initio* calculations, and experiments suggests that the explicit treatment of lattice vibrations is necessary to achieve qualitatively correct trends in $C_{44}$ vs. T. Based on empirical criteria used to assess ductility trends and previous hardness measurements (H ≈27 GPa), we propose the high-entropy (V,Nb,Ta,Mo,W)C ceramic as candidate carbide system with high toughness both at room and elevated temperature.

## Acknowledgements

All simulations were carried out using the resources provided by the Swedish National Infrastructure for Computing (SNIC) – partially funded by the Swedish Research Council through Grant Agreement N° VR-2015-04630 – on the Clusters located at the National Supercomputer Centre (NSC) in Linköping, the Center for High Performance Computing (PDC) in Stockholm, and at the High Performance Computing Center North (HPC2N) in Umeå, Sweden. We gratefully acknowledge financial support from the Competence Center Functional Nanoscale Materials (FunMat-II) (Vinnova Grant No. 2016–05156), the Swedish Research Council (VR) through Grant No. 2019–05600, the Swedish Government Strategic Research Area in Materials Science on Functional Materials at Linköping University (Faculty Grant SFO-Mat-LiU No. 2009-00971), and the Knut and Alice Wallenberg Foundation through Wallenberg Scholar project (Grant No. 2018.0194). D.G.S. gratefully acknowledges financial support from the Olle Engkvist Foundation. TH and KV would like to acknowledge support through the Office of Naval Research ONR-MURI (grant No. N00014-15-1-2863). Analysis of results of theoretical simulations was supported by RFBR Project N° 20-02-00178.

**Figures**

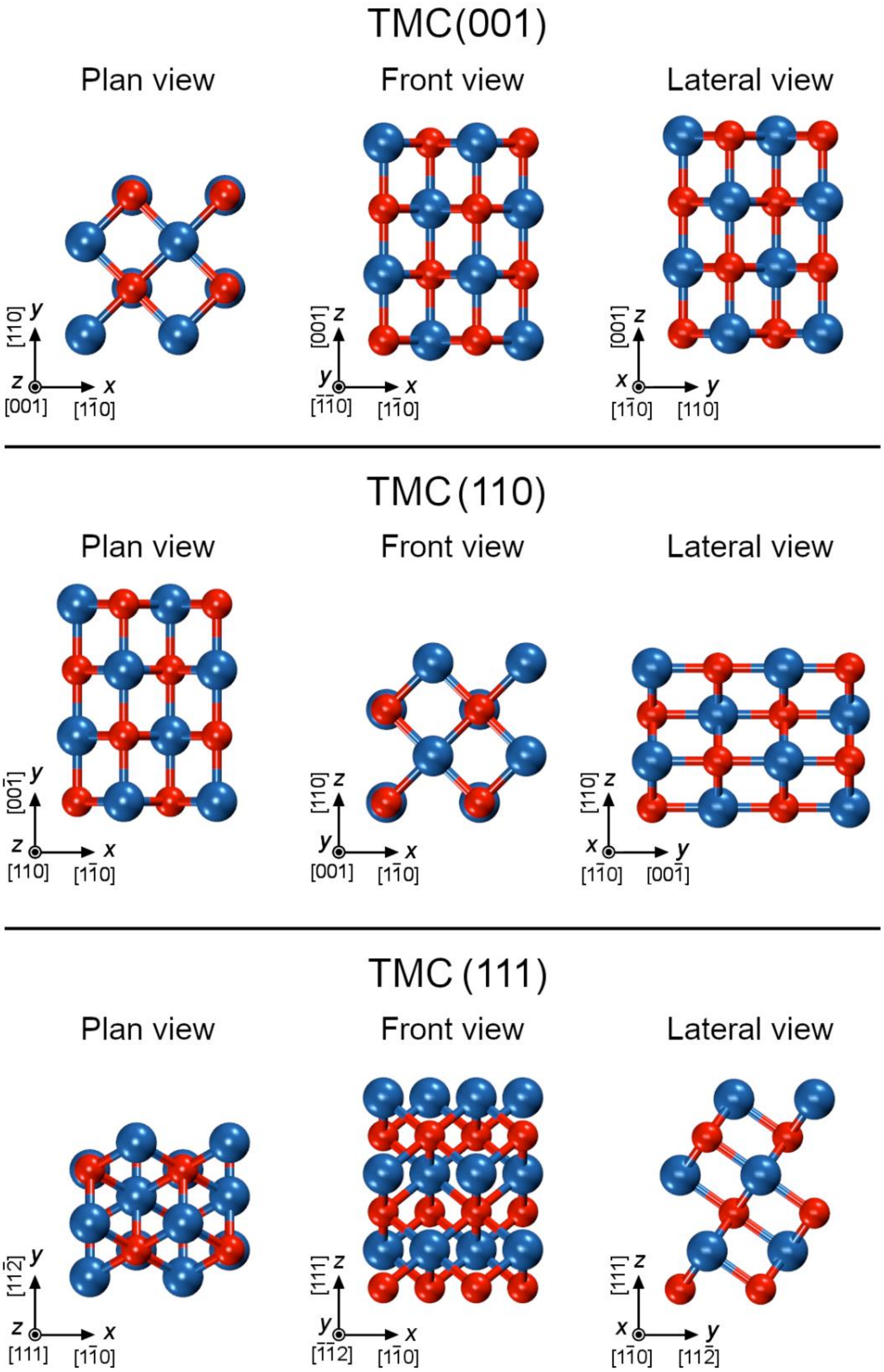

**Fig. 1**. Illustration of the building blocks used to generate supercell models with three different vertical (z) orientations, i.e., [001], [110], and [111]. Note that the actual supercells used in our AIMD simulations of tensile and shear deformation contain 576 atoms each (see figure 1 in [33]). Spheres of different colors indicate the metal and carbon sites of the B1 crystal structure. This figure is generated using the VMD software [89].

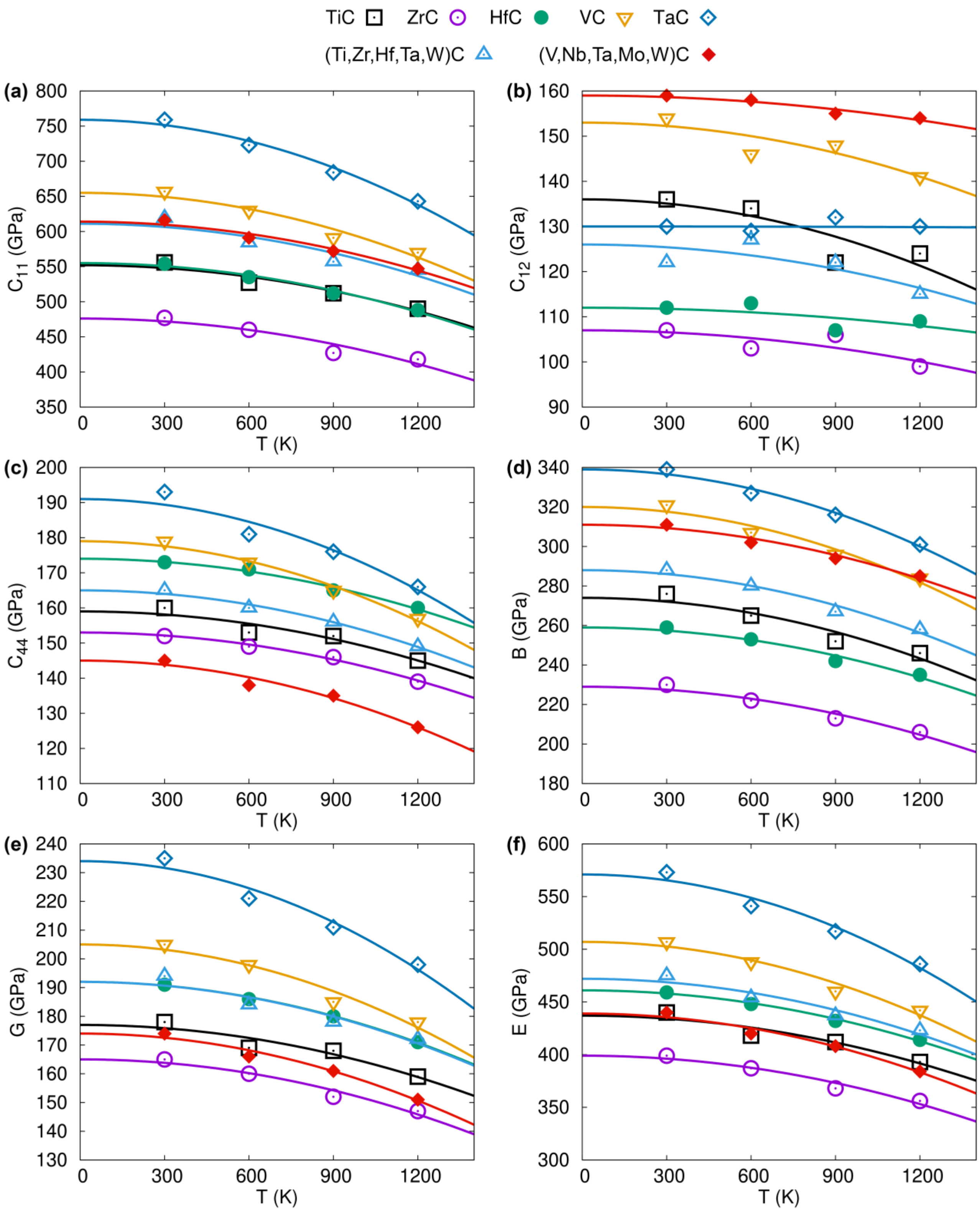

**Fig. 2**. **(a)** $C_{11}$, **(b)** $C_{12}$, **(c)** $C_{44}$ elastic constants, bulk **(d)** B, shear **(e)** G, and **(f)** Young E moduli (open and filled symbols) of B1 carbides as a function of temperature calculated via AIMD. The solid lines, which serve as guide to the eye, are 2$^{nd}$ order polynomials obtained by imposing zero slope at 0 K and least-squares differences with the elastic properties determined by AIMD. The statistical uncertainty on the calculated elastic properties is of the order of the pressure differences between symbols and corresponding lines.

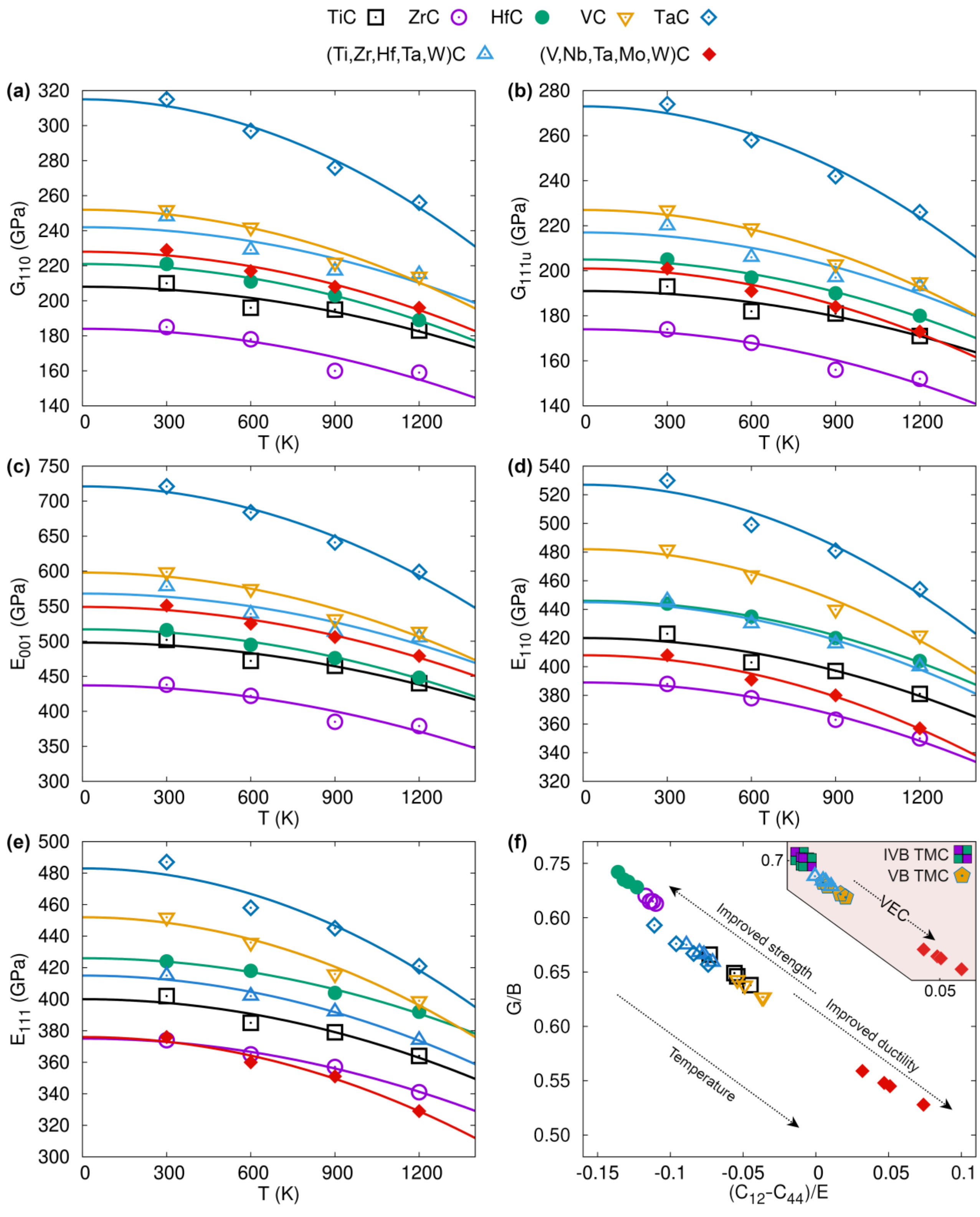

**Fig. 3**. Shear elastic stiffnesses **(a)** $G_{110}$ and **(b)** $G_{111u}$, directional Youngs moduli **(c)** $E_{001}$, **(d)** $E_{110}$, and **(e)** $E_{111}$ (open and filled symbols) and **(f)** ductility vs. strength maps based on phenomenological elastic-moduli-based indicators [7, 12] of B1 carbides as a function of temperature calculated via AIMD. The solid lines, which serve as guide to the eye, are 2$^{nd}$ order polynomials obtained by imposing zero slope at 0 K and least-squares differences with the elastic properties determined by AIMD. The statistical uncertainty on the calculated elastic properties is of the order of the difference in pressure (GPa) between a symbol and the corresponding fitted curve. In **(f)**, the empirical criteria suggest that temperature has the general effect of improving ductility while reducing strength. However, some carbides exhibit a non-monotonic behavior in such sense (see **Tables 3–9**).

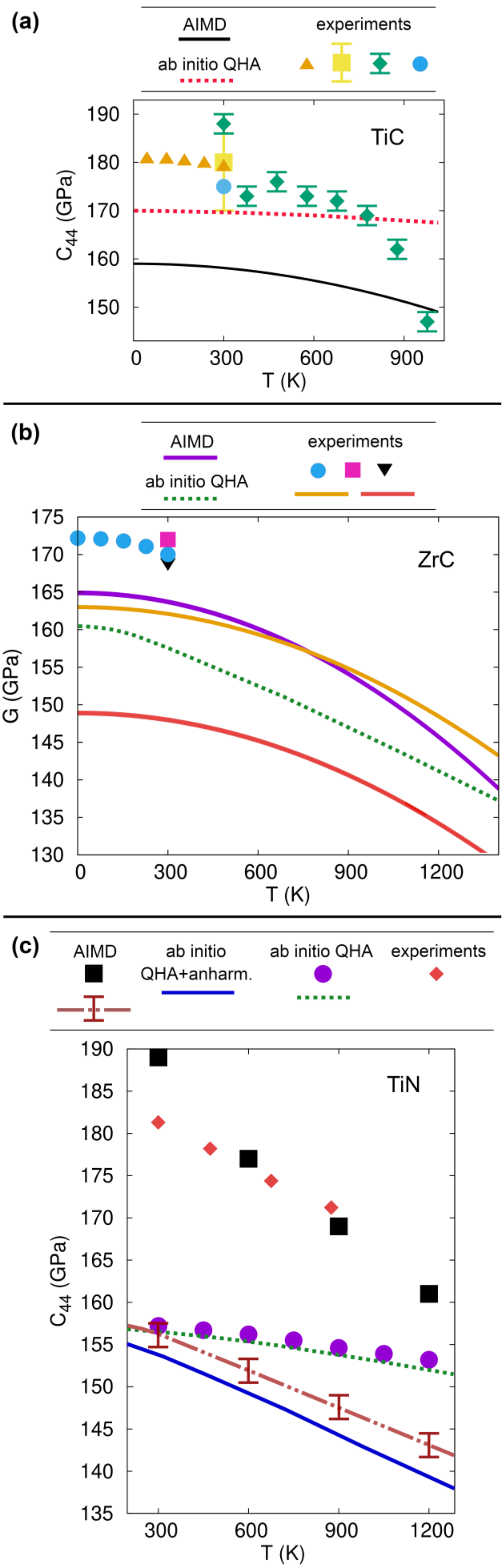

**Fig. 4**. Comparison between the trends in the shear elastic stiffness vs. T obtained via AIMD simulations (present work), previous *ab initio* QHA calculations [23] and experiments for different B1-structure ceramics. **(a)** $C_{44}$ vs. T in TiC. QHA results are from Ref. [23]; experimental results for $TiC_x$ ($x \leq 1$): green diamonds [90, 91], orange triangles [92], yellow square [93], blue circle [94]. **(b)** G vs. T in ZrC. QHA results are from Ref. [27]; experimental results for $ZrC_x$ ($x \leq 1$): pink square [36], black triangle [36], blue circles [92], orange solid line [67] {$B = E/[3(1\text{-}2\nu)]$ and $G = 3BE/(9B–E)$}, red solid line [66]. **(c)** $C_{44}$ vs. T in TiN. AIMD results of the present work are marked by black squares. Other AIMD results (red – ·· – line with error bars) are from Ref. [30]. *Ab initio* results that account for both thermal expansion and anharmonic effects (solid blue line) are from Ref. [28]. Results of static *ab initio* calculations: purple circles (QHA) [62] and green dotted line (volume scaling) [28]. The experimental results (red diamond) are from Ref. [95]. The figure shows that the temperature-induced decay in experimental shear elastic stiffness is better reproduced by AIMD simulations than by static *ab initio* calculations that account for thermal expansion but neglect lattice vibrations.

## Tables

| | $[\partial\sigma_{xx}/\partial\delta]_{\delta=0}$ | $[\partial\sigma_{yy}/\partial\delta]_{\delta=0}$ | $[\partial\sigma_{zz}/\partial\delta]_{\delta=0}$ |
|---|---|---|---|
| Tensile deformation | | | |
| ⟨001⟩ | $C_{12}$ | $C_{12}$ | $C_{11}$ |
| ⟨110⟩ | $(C_{11}+C_{12})/2-C_{44}$ | $C_{12}$ | $(C_{11}+C_{12})/2+C_{44}$ |
| ⟨111⟩ | $(C_{11}+2C_{12}-2C_{44})/3$ | $(C_{11}+2C_{12}-2C_{44})/3$ | $(C_{11}+2C_{12}+4C_{44})/3$ |
| | $[\partial\sigma_{xz}/\partial\delta]_{\delta=0}$ | $[\partial\sigma_{xy}/\partial\delta]_{\delta=0}$ | $[\partial\sigma_{yz}/\partial\delta]_{\delta=0}$ |
| Shear deformation | | | |
| {001}⟨1$\bar{1}$0⟩ | $C_{44}$ | 0 | 0 |
| {110}⟨1$\bar{1}$0⟩ | $(C_{11}-C_{12})/2$ | 0 | 0 |
| {111}⟨1$\bar{1}$0⟩ | $(C_{11}-C_{12}+C_{44})/3$ | $(C_{11}-C_{12}-2C_{44})/(3\sqrt{2})$ | 0 |

**Table 1.** Correspondences between stress variations $[\partial\sigma_{ij}/\partial\delta]_{\delta=0}$ obtained during tensile or shear deformation for vanishingly small strains and second-order elastic constants of a cubic crystal. All $\sigma_{ij}$ values equal 0 mean that the derivative of that $\sigma_{ij}$ vs. strain component is equal zero for $\delta=0$. We note that the supercell volume and shape remain constant at each strain step of AIMD/NVT simulations. Thus, the materials' elastic resistances to ⟨001⟩, ⟨110⟩, and ⟨111⟩ elongation *do not* correspond to the directional Young's moduli $E_{001}$, $E_{110}$, $E_{111}$. A direct evaluation of $E_{hkl}$ during tensile strain would require lateral supercell relaxation.

| T (K) | Lattice parameter (Å) | | | | | | |
|---|---|---|---|---|---|---|---|
| | TiC | ZrC | HfC | VC | TaC | (Ti,Zr,Hf,Ta,W)C | (V,Nb,Ta,Mo,W)C |
| 0 | {4.331[b],<br>4.34[w], 4.261–4.348[y],<br>4.297[f]} | {4.718[b],<br>4.73[w],<br>4.685[f]} | {4.642[b],<br>4.65[w],<br>4.608[f], 4.610–4.726[Ω]} | {4.150[b], 4.16[w],<br>4.125[f]} | {4.471[b], 4.48[w],<br>4.446[f], 4.435–<br>4.555[Ω]} | | |
| 300 | **4.307**<br>[4.318[q], 4.328[k],<br>4.327[j], 4.330[t1]] | **4.677**<br>[4.683[q], 4.700[k],<br>4.696[j]] | **4.616**<br>[4.637[k]<br>4.641[j]] | **4.133**<br>[4.164[k]] | **4.450**<br>[4.454[q], 4.455[k],<br>4.454[p*]] | **4.477**<br>[4.487[c]] | **4.370**<br>[4.340[c]] |
| 600 | **4.317** | **4.687** | **4.625** | **4.143** | **4.460** | **4.487** | **4.379** |
| 900 | **4.327** | **4.698** | **4.634** | **4.154** | **4.469** | **4.498** | **4.389** |
| 1200 | **4.338**<br>[4.345[q]] | **4.709**<br>[4.715[q]] | **4.645** | **4.166** | **4.480**<br>[4.483[q]] | **4.508** | **4.400** |

| $\alpha$ ($10^{-6}\cdot K^{-1}$) | | | | | | |
|---|---|---|---|---|---|---|
| TiC | ZrC | HfC | VC | TaC | (Ti,Zr,Hf,Ta,W)C | (V,Nb,Ta,Mo,W)C |
| **8.0**<br>[7.6±0.2[q], 8.5[t], 9.5±0.3[j], 8.5[k]]<br>{6.4[y]} | **7.6**<br>[7.0±0.2[q],<br>7.6±0.2[j], 7.5[k], 7.5[e]] | **7.0**<br>[7.3±0.2[j], 6.1[k]] | **8.9**<br>[7.2[k]] | **7.5**<br>[6.6±0.2[q], 6.3[k]] | **7.7** | **7.6** |

**Table 2**.* Structural parameters of TMC determined via AIMD/NPT simulations as a function of temperature. The average linear thermal expansion $\alpha$ is obtained for AIMD lattice parameters between 300 and 1200 K.

[Ω] [96] (see supplemental material), *ab initio* calculations based on PBE and PBE with different semi-local and non-local van der Waals approximations.
[f] [60], *ab initio* calculations, PBEsol functional.
[b] [61], *ab initio* calculations.
[c] [37], experiments for (Ti,Zr,Hf,Ta,W)C and (V,Nb,Ta,Mo,W)C.
[j] [97], experiments between 300 and 3000 K for $TiC_{0.73}$, $ZrC_{0.75}$, and $HfC_{0.88}$.
[k] [98], experiments for $VC_{0.88}$.
[p*] [99], experiments for $TaC_{0.99}$; data refer to samples with 6% porosity.
[q] [100], experiments between 300 and ≈1200 K for $TiC_{0.92}$, $ZrC_{0.96}$, and $TaC_{0.98}$.
[t] [90], experiment: $TiC_{0.97}$. The average linear thermal expansion has been measured between 300 and 1600 K.
[w] [101], *ab initio* calculations.
[e] [65], experiments for $ZrC_{0.96}$. and $ZrC_{0.92}$.
[y] [23], *ab initio* calculations: the range of lattice parameters includes GGA and LDA results from Ref. [23] and *ab initio* studies referenced therein. The coefficient of thermal expansion (GGA approximation) is obtained by averaging over the values computed between 300 and 1200 K.
[t1] [91], experiments for $TiC_{0.97}$. Specimen with a maximum porosity of 1.3%.

| T (K) | $C_{11}$ (GPa) | $C_{12}$ (GPa) | $C_{44}$ (GPa) | B (GPa) | $G_{110}$ (GPa) | $G_{111u}$ (GPa) | $G_{111r}$ (GPa) | $E_{001}$ (GPa) | $E_{110}$ (GPa) | $E_{111}$ (GPa) | $G_V$ (GPa) | $G_R$ (GPa) | $G_H$ (GPa) | $E_V$ (GPa) | $E_R$ (GPa) | $E_H$ (GPa) | A (%) | ν (%) | $G_H/B$ | $(C_{44}–C_{12})/E_H$ |
|---|---|---|---|---|---|---|---|---|---|---|---|---|---|---|---|---|---|---|---|---|
| 0 | [522[a]] {546[b], 550[d], 509[g], 530[w], 470–606[y], 557[f]} | [106[a]] {133[b], 130[d], 146[g], 130[w], 97–130[y], 124[f]} | [181[a]] {162[b], 165[d], 168[g], 163[w], 167–230[y], 163[f]} | [245[a]] {270[b], 270[d], 267[g], 263[w], 268[f], 251[m]} | {200[n]} | | | | | | | | {182[d], 183[f], 181[m]} | | | {445[d], 431[g], 448[f], 438[m]} | | {23[d], 21[g], 22[f]} | {0.71[g], 0.674[d]} | |
| 300 | **556** [540±30[i], 489[t], 477[t1], 500[h], 515[a]] {490[y]} | **136** [110±30[i], 168[t], 98[t1], 113[h], 106[a]] {118[y]} | **160** [180±10[i], 187[t], 190[t1], 175[h], 179[a]] {170[y]} | **276** [275[t], 224[t1], 242[h], 242[a], 233±14[r], 241[m], 255[m]] | **210** | **193** | **190** | **502** | **423** | **402** | **180** [184±11[r], 186[m], 207[m], 187[t1]] | **177** [184±11[r], 186[m], 207[m], 187[t1]] | **178** [184±11[r], 186[m], 207[m], 187[t1]] | **443** [444[t1] 450[k], 436±26[r], 447–451[m], 489±13[m]] | **437** [444[t1] 450[k], 436±26[r], 447–451[m], 489±13[m]] | **440** [444[t1] 450[k], 436±26[r], 447–451[m], 489±13[m]] | **76** | **23** [19±1[r]] | **0.646** | **-0.054** |
| 600 | **527** [450[t], 439[t1]] {480[y]} | **134** [135[t], 90[t1]] {114[y]} | **153** [172[t], 175[t1]] {169[y]} | **265** [240[t], 206[t1]] | **196** | **182** | **179** | **472** | **403** | **385** | **170** [173[t1]] | **168** [173[t1]] | **169** [173[t1]] | **421** [409[t1]] | **416** [409[t1]] | **418** [409[t1]] | **78** | **24** | **0.638** | **-0.045** |
| 900 | **512** [423[t], 413[t1]] {469[y]} | **122** [110[t], 85[t1]] {111[y]} | **152** [162[t], 164[t1]] {168[y]} | **252** [214[t], 194[t1]] | **195** | **181** | **178** | **465** | **397** | **379** | **169** [163[t1]] | **167** [163[t1]] | **168** [163[t1]] | **414** [384[t1]] | **409** [384[t1]] | **412** [384[t1]] | **78** | **23** | **0.666** | **-0.073** |
| 1200 | **490** {457[y]} | **124** {107[y]} | **145** {166[y]} | **246** | **183** | **171** | **169** | **440** | **381** | **364** | **161** | **158** | **159** | **395** [379[u]] | **391** [379[u]] | **393** [379[u]] | **79** | **23** | **0.649** | **-0.056** |

**Table 3**.* Temperature-dependent elastic properties of B1 TiC.

[f] [60], *ab initio* calculations (PBEsol functional).
[b] [61], *ab initio* calculations.
[d] [39], *ab initio* calculations. Poisson's ratio recalculated using $E_H$ and B from ref. [39] in formula $\nu = [1–E_H/(3B)]/2$.
[w] [101], *ab initio* calculations.
[y] [23], *ab initio* calculations based on quasiharmonic approximation. The range of elastic constant values includes GGA and LDA results from Ref. [23] and *ab initio* studies referenced therein.
[g] [16] (see supplemental material), *ab initio* calculations.
[n] [102], *ab initio* calculations.
[a] [92], experiments for $ZrC_{0.94}$ and $TiC_{0.91}$ from 8 K to room temperature.
[h] [94], experiments for TiC; stoichiometry not specified.
[i] [93], experiments for TiC.
[j] [97], experiments between 300 and 3000 K for $TiC_{0.73}$, $ZrC_{0.75}$, and $HfC_{0.88}$.
[q] [100], experiments between 300 and ≈1200 K for $TiC_{0.92}$, $ZrC_{0.96}$, and $TaC_{0.98}$.
[r] [103], experiments for $TiC_{0.98}$ and $TaC_{0.98}$ (see porosity corrected-values in table 3).
[t] [90], experiments for $TiC_{0.97}$. Evaluation of elastic properties assumes isotropic specimen and temperature-independent Poisson's ratio.
[t1] [91], experiments for $TiC_{0.97}$. Evaluation of elastic properties assumes isotropic specimen (which has a maximum porosity of 1.3%) and temperature-independent Poisson's ratio ν = 0.17.
[u] [104], experiments for TiC at 1273 K.
[m] [36], *ab initio* calculations and experiments + experimental references therein.

| T (K) | $C_{11}$ (GPa) | $C_{12}$ (GPa) | $C_{44}$ (GPa) | B (GPa) | $G_{110}$ (GPa) | $G_{111u}$ (GPa) | $G_{111r}$ (GPa) | $E_{001}$ (GPa) | $E_{110}$ (GPa) | $E_{111}$ (GPa) | $G_V$ (GPa) | $G_R$ (GPa) | $G_H$ (GPa) | $E_V$ (GPa) | $E_R$ (GPa) | $E_H$ (GPa) | A (%) | ν (%) | $G_H$/B | ($C_{44}$–$C_{12}$)/$E_H$ |
|---|---|---|---|---|---|---|---|---|---|---|---|---|---|---|---|---|---|---|---|---|
| 0 | [480[a]] {490[b], 468[g], 474[w], 482[f], 450[o]} | [99[a]] {112[b], 123[g], 115[w], 113[f], 103[o]} | [161[a]] {147[b], 152[g], 144[w], 146[f], 153[o]} | [226[a]] {238[b], 238[g], 235[w], 236[f], 221[m]} | | | | | | | | | {160[f], 157[m]} | | | {444[g], 392[f], 381[m]} | | {19[g], 22[f]} | {0.77[g]} | |
| 300 | **477** [472[a]] {444[o], 435[o]} | **107** [99[a]] {103±1[o]} | **152** [159[a]] {153[o]} | **230** [223[a], 241[e], 220[m], 216[m]] | **185** | **174** | **173** | **438** | **388** | **374** | **165** [170[a], 148[s], 172[m], 169[m], 163[e]] | **164** [170[a], 148[s], 172[m], 169[m], 163[e]] | **165** [170[a], 148[s], 172[m], 169[m], 163[e]] | **400** [406[a], 390–426[e], 346[s], 350[k], 385–406[m], 402±13[m]] | **397** [406[a], 390–426[e], 346[s], 350[k], 385–406[m], 402±13[m]] | **399** [406[a], 390–426[e], 346[s], 350[k], 385–406[m], 402±13[m]] | **82** | **21** [20[a], 23[e]] | **0.715** | **-0.114** |
| 600 | **460** {432±2[o], 414±2[o]} | **103** {103±7[o]} | **149** {151[o]} | **222** [239[e]] | **178** | **168** | **167** | **422** | **378** | **365** | **161** [146[s], 159[e]] | **159** [146[s], 159[e]] | **160** [146[s], 159[e]] | **388** [385–411[e], 341[s]] | **386** [385–411[e], 341[s]] | **387** [385–411[e], 341[s]] | **83** | **21** [23[e]] | **0.720** | **-0.117** |
| 900 | **427** {419±4[o], 391±4[o]} | **106** {101±9[o]} | **147** {150[o]} | **213** [235[e]] | **161** | **156** | **156** | **385** | **364** | **357** | **152** [141[s], 154[e]] | **152** [141[s], 154[e]] | **152** [141[s], 154[e]] | **369** [375–394[e], 333[s]] | **368** [375–394[e], 333[s]] | **368** [375–394[e], 333[s]] | **91** | **21** [24[e]] | **0.715** | **-0.111** |
| 1200 | **418** {405±6[o], 367±6[o]} | **99** {102±9[o]} | **139** {149[o]} | **205** [232[e]] | **160** | **153** | **152** | **380** | **350** | **341** | **147** [134[s], 149[e]] | **147** [134[s], 149[e]] | **147** [134[s], 149[e]] | **357** [363–383[e], 321[s]] | **355** [363–383[e], 321[s]] | **356** [363–383[e], 321[s]] | **87** | **21** [24[e]] | **0.716** | **-0.113** |

**Table 4**.* Temperature-dependent elastic properties of B1 ZrC.

[f] [60], *ab initio* calculations (PBEsol functional).
[b] [61], *ab initio* calculations.
[w] [101], *ab initio* calculations.
[g] [16] (see supplemental material), *ab initio* calculations.
[s] [66], experiments $ZrC_{0.86}$. Carbon content of 10 weight% corresponds to 86% carbon occupancy of the anion sublattice in B1 ZrC.
[e] [65, 67], experiments for $ZrC_{0.96}$. and $ZrC_{0.92}$. B estimated as E/[3(1-2$\nu$)] and G = 3BE/(9B-E)
[a] [92], experiments for $ZrC_{0.94}$ and $TiC_{0.91}$ from 8 K to room temperature.
[j] [97], experiments between 300 and 3000 K for $TiC_{0.73}$, $ZrC_{0.75}$, and $HfC_{0.88}$.
[q] [100], experiments between 300 and ≈1200 K for $TiC_{0.92}$, $ZrC_{0.96}$, and $TaC_{0.98}$.
[m] [36], *ab initio* calculations and experiments + experimental references therein.
[o] [27], *ab initio* calculations based on quasiharmonic and quasistatic approximations.

| T (K) | $C_{11}$ (GPa) | $C_{12}$ (GPa) | $C_{44}$ (GPa) | B (GPa) | $G_{110}$ (GPa) | $G_{111u}$ (GPa) | $G_{111r}$ (GPa) | $E_{001}$ (GPa) | $E_{110}$ (GPa) | $E_{111}$ (GPa) | $G_V$ (GPa) | $G_R$ (GPa) | $G_H$ (GPa) | $E_V$ (GPa) | $E_R$ (GPa) | $E_H$ (GPa) | A (%) | ν (%) | $G_H/B$ | $(C_{44}–C_{12})/E_H$ |
|---|---|---|---|---|---|---|---|---|---|---|---|---|---|---|---|---|---|---|---|---|
| 0 | {520[g], 525[w], 551[f], 540[b], 501[o]} | {133[g], 118[w], 105[f], 112[b], 102[o]} | {172[g], 164[w], 175[f], 171[b], 177[o]} | {262[g], 254[w], 254[f], 255[b], 239[m]} | {236[n]} | | | | | | | | {192[f], 186[m]} | | | {395[g], 461[f], 443[m]} | | {20[g], 20[f]} | {0.74[g]} | |
| 300 | **554** {496[o], 489[o]} | **112** {102[o]} | **173** {176±1[o]} | **259** [243[z], 223[m], 241[m]] | **221** | **205** | **202** | **516** | **444** | **424** | **192** [195[z], 179–193[m], 181[m]] | **189** [195[z], 179–193[m], 181[m]] | **191** [195[z], 179–193[m], 181[m]] | **462** [462[z], 420[k], 316–461[m], 428±32[m]] | **457** [462[z], 420[k], 316–461[m], 428±32[m]] | **459** [462[z], 420[k], 316–461[m], 428±32[m]] | **78** | **20** [19[z]] | **0.735** | **-0.132** |
| 600 | **535** {483±3[o], 464±3[o]} | **113** {103±6[o]} | **171** {174±2[o]} | **253** | **211** | **197** | **195** | **495** | **435** | **418** | **187** | **185** | **186** | **450** | **446** | **448** | **81** | **20** | **0.733** | **-0.129** |
| 900 | **512** {467±5[o], 441±5[o]} | **107** {103±8[o]} | **165** {171±3[o]} | **242** | **203** | **190** | **189** | **476** | **420** | **404** | **180** | **179** | **180** | **433** | **430** | **432** | **82** | **20** | **0.742** | **-0.136** |
| 1200 | **488** {455±6[o], 416±6[o]} | **109** {103±9[o]} | **160** {170±4[o]} | **235** | **189** | **180** | **179** | **448** | **404** | **392** | **172** | **171** | **171** | **415** | **412** | **414** | **85** | **21** | **0.728** | **-0.123** |

**Table 5**.* Temperature-dependent elastic properties of B1 HfC.

[f] [60], *ab initio* calculations (PBEsol functional).
[b] [61], *ab initio* calculations.
[w] [101], *ab initio* calculations.
[g] [16] (see supplemental material), *ab initio* calculations.
[n] [102], *ab initio* calculations, GGA functional.
[z] [105], experiments for $HfC_{0.97}$ and $TaC_{0.99}$.
[j] [97], experiments between 300 and 3000 K for $TiC_{0.73}$, $ZrC_{0.75}$, and $HfC_{0.88}$.
[z] [105], experiments for $HfC_{0.97}$ and $TaC_{0.99}$.
[m] [36], *ab initio* calculations and experiments + experimental references therein.
[o] [27], *ab initio* calculations based on quasiharmonic and quasistatic approximations.

| T (K) | $C_{11}$ (GPa) | $C_{12}$ (GPa) | $C_{44}$ (GPa) | B (GPa) | Δ (GPa) | $G_{110}$ (GPa) | $G_{111u}$ (GPa) | $G_{111r}$ (GPa) | $E_{001}$ (GPa) | $E_{110}$ (GPa) | $E_{111}$ (GPa) | $G_V$ (GPa) | $G_R$ (GPa) | $G_H$ (GPa) | $E_V$ (GPa) | $E_R$ (GPa) | $E_H$ (GPa) | A (%) | ν (%) | $G_H/B$ | $(C_{44}–C_{12})/E_H$ |
|---|---|---|---|---|---|---|---|---|---|---|---|---|---|---|---|---|---|---|---|---|---|
| 0 | {669[b], 648[g], 652[w], 706[d], 705[f]} | {152[b], 170[g], 141[w], 124[d], 141[f]} | {186[b], 189[g], 188[w], 185[d], 190[f]} | {332[b], 329[g], 311[w], 318[d], 329[f], 283[m]} | | | | | | | | | | {222[d], 223[f], 199[m]} | | | {520[g], 523[d], 546[f], 484[m]} | | {21[g], 23[d], 22[f]} | {0.68[g], 0.674[d]} | |
| 300 | **657** [366–527[v]] | **154** [84–292[v]] | **179** [136–192[v]] | **321** [250[m], 389[m]] | **44** | **252** | **227** | **222** | **599** | **482** | **452** | **208** [157[m], 196[m]] | **202** [157[m], 196[m]] | **205** [157[m], 196[m]] | **513** [430[k], 268–420[m], 465±13[m]] | **501** [430[k], 268–420[m], 465±13[m]] | **507** [430[k], 268–420[m], 465±13[m]] | **71** | **24** | **0.638** | **-0.049** |
| 600 | **630** | **146** | **173** | **307** | **55** | **242** | **219** | **213** | **575** | **464** | **436** | **200** | **195** | **198** | **494** | **483** | **488** | **71** | **23** | **0.643** | **-0.054** |
| 900 | **595** | **148** | **161** | **297** | **49** | **224** | **203** | **198** | **536** | **434** | **408** | **186** | **181** | **183** | **461** | **451** | **456** | **72** | **24** | **0.618** | **-0.028** |
| 1200 | **533** | **132** | **156** | **265** | **76** | **201** | **186** | **183** | **481** | **411** | **392** | **174** | **172** | **173** | **429** | **424** | **426** | **78** | **23** | **0.651** | **-0.058** |

**Table 6**.* Temperature-dependent elastic properties of B1 VC.

[f] [60], *ab initio* calculations (PBEsol functional).
[b] [61], *ab initio* calculations.
[d] [39], *ab initio* calculations. Poisson's ratio recalculated using formula $\nu = [1–E_H/(3B)]/2$.
[w] [101], *ab initio* calculations.
[g] [16] (see supplemental material), *ab initio* calculations.
[k] [98], experiments for $VC_{0.88}$.
[m] [36], *ab initio* calculations and experiments + experimental references therein.
[v] [106], experiments $VC_x$ ($0.75 \leq x \leq 0.88$).

| T (K) | $C_{11}$ (GPa) | $C_{12}$ (GPa) | $C_{44}$ (GPa) | B (GPa) | $G_{110}$ (GPa) | $G_{111u}$ (GPa) | $G_{111r}$ (GPa) | $E_{001}$ (GPa) | $E_{110}$ (GPa) | $E_{111}$ (GPa) | $G_V$ (GPa) | $G_R$ (GPa) | $G_H$ (GPa) | $E_V$ (GPa) | $E_R$ (GPa) | $E_H$ (GPa) | A (%) | ν (%) | $G_H$/B | ($C_{44}$–$C_{12}$)/$E_H$ |
|---|---|---|---|---|---|---|---|---|---|---|---|---|---|---|---|---|---|---|---|---|
| 0 | {737[b], 722[g], 685[w], 778[f]} | {141[b], 157[g], 163[w], 128[f]} | {175[b], 176[g], 158[w], 181[f]} | {340[b], 345[g], 337[w], 344[f], 326[m]} | | | | | | | | | {229[f], 213[m]} | | | {553[g], 563[f], 525[m]} | | {22[g], 23[f]} | {0.68[g]} | |
| 300 | **759** | **130** | **193** | **339** [331[p], 287[p*], 345[z], 332±39[r], 248–343[m], 219[m]] | **315** | **274** | **260** | **721** | **530** | **487** | **242** [227[p], 202[p*], 217[z], 234±27[r], 215–227[m], 184[m]] | **228** [227[p], 202[p*], 217[z], 234±27[r], 215–227[m], 184[m]] | **235** [227[p], 202[p*], 217[z], 234±27[r], 215–227[m], 184[m]] | **586** [561[p], 490[p*], 523[l], 496[x], 290[k], 538[z], 567±68[r], 241–722[m], 431±44[m]] | **560** [561[p], 490[p*], 523[l], 496[x], 290[k], 538[z], 567±68[r], 241–722[m], 431±44[m]] | **573** [561[p], 490[p*], 523[l], 496[x], 290[k], 538[z], 567±68[r], 241–722[m], 431±44[m]] | **61** | **22** [21[p], 24[z], 21±2[r], 21[x]] | **0.693** | **-0.111** |
| 600 | **723** | **129** | **181** | **327** [275[p*]] | **297** | **258** | **245** | **684** | **499** | **458** | **227** [187[p*]] | **214** [187[p*]] | **221** [187[p*]] | **554** [464[p*], 504[l], 484[x]] | **528** [464[p*], 504[l], 484[x]] | **541** [464[p*], 504[l], 484[x]] | **61** | **22** [22[p], 22[x]] | **0.676** | **-0.096** |
| 900 | **684** | **132** | **176** | **316** [273[p*]] | **276** | **242** | **232** | **641** | **481** | **445** | **216** [182[p*]] | **206** [182[p*]] | **211** [182[p*]] | **527** [448[p*], 492[l], 473[x]] | **507** [448[p*], 492[l], 473[x]] | **517** [448[p*], 492[l], 473[x]] | **64** | **23** [23[p], 22[x]] | **0.667** | **-0.084** |
| 1200 | **643** | **130** | **166** | **301** [268[p*]] | **256** | **226** | **217** | **599** | **454** | **421** | **202** [176[p*]] | **193** [176[p*]] | **198** [176[p*]] | **495** [434[p*], 480[l], 462[x]] | **477** [434[p*], 480[l], 462[x]] | **486** [434[p*], 480[l], 462[x]] | **65** | **23** [23[p], 22[x]] | **0.657** | **-0.074** |

**Table 7**.* Temperature-dependent elastic properties of B1 TaC.

[f] [60], *ab initio* calculations (PBEsol functional).
[b] [61], *ab initio* calculations.
[w] [101], *ab initio* calculations.
[g] [16] (see supplemental material), *ab initio* calculations.
[l] [107] experiments for $TaC_{0.99}$.
[x] [108] experiments for $TaC_{0.95}$ with 3% porosity.
[z] [105], experiments for $HfC_{0.97}$ and $TaC_{0.99}$.
[p] [99], experiments for $TaC_{0.99}$; data extrapolated to zero porosity.
[p*] [99], experiments for $TaC_{0.99}$; data refer to samples with 6% porosity.
[q] [100], experiments between 300 and ≈1200 K for $TiC_{0.92}$, $ZrC_{0.96}$, and $TaC_{0.98}$.
[r] [103], experiments for $TiC_{0.98}$ and $TaC_{0.98}$ (see porosity corrected-values in table 3).

[m] [36], *ab initio* calculations and experiments + experimental references therein.

| T (K) | $C_{11}$ (GPa) | $C_{12}$ (GPa) | $C_{44}$ (GPa) | B (GPa) | $G_{110}$ (GPa) | $G_{111u}$ (GPa) | $G_{111r}$ (GPa) | $E_{001}$ (GPa) | $E_{110}$ (GPa) | $E_{111}$ (GPa) | $G_V$ (GPa) | $G_R$ (GPa) | $G_H$ (GPa) | $E_V$ (GPa) | $E_R$ (GPa) | $E_H$ (GPa) | A (%) | ν (%) | $G_H/B$ | $(C_{44}–C_{12})/E_H$ |
|---|---|---|---|---|---|---|---|---|---|---|---|---|---|---|---|---|---|---|---|---|
| 0 | | | | {274[m]} | | | | | | | | | {191[m]} | | | {466[m]} | | {22[m]} | | |
| 300 | **619** | **122** | **165** | **288** (273±17) [246[m]] | **248** | **220** | **212** | **578** | **446** | **415** | **198** (193±4) [200[m]] | **190** (193±4) [200[m]] | **194** (193±4) [200[m]] | **483** (468±3) [473±26[m]] | **468** (468±3) [473±26[m]] | **475** (468±3) [473±26[m]] | **66** | **22** (21±2) [18±2[m]] | **0.675** | **-0.089** |
| 600 | **584** | **127** | **160** | **280** | **229** | **206** | **200** | **539** | **430** | **402** | **187** | **182** | **184** | **459** | **448** | **454** | **70** | **23** | **0.659** | **-0.071** |
| 900 | **557** | **122** | **156** | **267** | **217** | **197** | **192** | **513** | **416** | **392** | **180** | **176** | **178** | **442** | **433** | **437** | **72** | **23** | **0.666** | **-0.077** |
| 1200 | **545** | **115** | **149** | **258** | **215** | **193** | **187** | **505** | **400** | **374** | **175** | **170** | **172** | **429** | **418** | **423** | **69** | **23** | **0.668** | **-0.080** |

**Table 8**.* Temperature-dependent elastic properties of B1 (Ti,Zr,Hf,Ta,W)C. The results of our present acoustic wave measurements are marked in bold and enclosed in ( ) brackets.

[m] [36], *ab initio* calculations and experiments.

| T (K) | $C_{11}$ (GPa) | $C_{12}$ (GPa) | $C_{44}$ (GPa) | B (GPa) | $G_{110}$ (GPa) | $G_{111u}$ (GPa) | $G_{111r}$ (GPa) | $E_{001}$ (GPa) | $E_{110}$ (GPa) | $E_{111}$ (GPa) | $G_V$ (GPa) | $G_R$ (GPa) | $G_H$ (GPa) | $E_V$ (GPa) | $E_R$ (GPa) | $E_H$ (GPa) | A (%) | ν (%) | $G_H/B$ | $(C_{44}–C_{12})/E_H$ |
|---|---|---|---|---|---|---|---|---|---|---|---|---|---|---|---|---|---|---|---|---|
| 0 | | | | {312[m]} | | | | | | | | | {183[m]} | | | {460[m]} | | {25[m]} | | |
| 300 | **616** | **159** | **145** | **311** (269±12) [278[m]] | **229** | **201** | **192** | **551** | **408** | **376** | **178** (173±4) [226[m]] | **170** (173±4) [226[m]] | **174** (173±4) [226[m]] | **449** (426±5) [533±32[m]] | **431** (426±5) [533±32[m]] | **440** (426±5) [533±32[m]] | **63** | **26** (24±1) [18±2[m]] | **0.559** | **0.032** |
| 600 | **591** | **158** | **138** | **302** | **217** | **191** | **182** | **525** | **391** | **360** | **170** | **162** | **166** | **429** | **412** | **420** | **64** | **27** | **0.548** | **0.047** |
| 900 | **572** | **155** | **135** | **294** | **208** | **184** | **176** | **506** | **380** | **351** | **164** | **157** | **161** | **415** | **400** | **408** | **65** | **27** | **0.545** | **0.051** |
| 1200 | **547** | **154** | **126** | **285** | **196** | **173** | **166** | **479** | **357** | **329** | **154** | **147** | **151** | **392** | **377** | **384** | **64** | **28** | **0.528** | **0.074** |

**Table 9**.* Temperature-dependent elastic properties of B1 (V,Nb,Ta,Mo,W)C. The results of our present acoustic wave measurements are marked in bold and enclosed in ( ) parentheses.

[m] [36], *ab initio* calculations and experiments.

*Elastic properties evaluated in the present work by AIMD simulations are in bold. Previous *ab initio* and experimental results are reported in curly {} and square [] brackets, respectively. Several elastic properties have been extracted from plots of previous works using the open software Plot Digitizer [109].